\documentclass[twocolumn,showpacs,preprintnumbers,amsmath,amssymb]{revtex4}
\usepackage{bm}
\usepackage{graphicx}
\usepackage{dcolumn}

\def\longrightharpoonup{\relbar\joinrel\rightharpoonup}
\def\longleftharpoondown{\leftharpoondown\joinrel\relbar}
\def\longrightleftharpoons{
  \mathop{
    \vcenter{
      \hbox{
      \ooalign{
        \raise1pt\hbox{$\longrightharpoonup\joinrel$}\crcr
	  \lower1pt\hbox{$\longleftharpoondown\joinrel$}
	  }
      }
    }
  }
}
\newcommand{\reversible}[2]{\displaystyle
  \mathrel{\longrightleftharpoons^{#1\mathstrut}_{#2}}}

\newcommand{\dd}{\ensuremath{\mathrm{d}}}

\newcommand{\bkappa}{\bm{\kappa}}

\newcommand{\beff}{\ensuremath{\mathbf{f}}}

\newcommand{\bg}{\ensuremath{\mathbf{g}}}
\newcommand{\bolx}{\overline{\bx}}
\newcommand{\ovx}{\overline{x}}

\newcommand{\be}{\begin{equation}}
\newcommand{\ee}{\end{equation}}
\newcommand{\bd}{\begin{displaymath}}
\newcommand{\ed}{\end{displaymath}}
\newcommand{\BE}{\begin{eqnarray}}
\newcommand{\EE}{\end{eqnarray}}

\newcommand{\bq}{\ensuremath{\mathbf{q}}}

\newcommand{\bv}{\ensuremath{\mathbf{v}}}

\newcommand{\bx}{\ensuremath{\mathbf{x}}}

\newcommand{\bn}{\ensuremath{\mathbf{n}}}
\newcommand{\bolde}{\ensuremath{\mathbf{e}}}

\newcommand{\bA}{\ensuremath{\mathbf{A}}}

\newcommand{\bp}{\ensuremath{\mathbf{p}}}

\newcommand{\bxi}{\bm{\xi}}
\newcommand{\olx}{\overline{\mathbf{x}}}

\newcommand{\avg}[1]{\left\langle{#1}\right\rangle}

\begin{document}

\preprint{}
\title{Limit cycles, complex Floquet multipliers and intrinsic noise}
\author{Richard P. Boland}
\email{richard.p.boland@postgrad.man.ac.uk}
\author{Tobias Galla}
\email{tobias.galla@manchester.ac.uk}
\author{Alan J. McKane}
\email{alan.mckane@manchester.ac.uk}
\affiliation{Theoretical Physics, School of Physics and Astronomy, University 
of Manchester, Manchester M13 9PL, United Kingdom}

\date{\today}

\begin{abstract}
We study the effects of intrinsic noise on chemical reaction systems, which 
in the deterministic limit approach a limit cycle in an oscillatory manner. 
Previous studies of systems with an oscillatory approach to a fixed point have
shown that the noise can transform the oscillatory decay into sustained 
coherent oscillations with a large amplitude. We show that a similar effect 
occurs when the stable attractors are limit cycles. We compute the correlation 
functions and spectral properties of the fluctuations in suitably co-moving 
Frenet frames for several model systems including driven and coupled 
Brusselators, and the Willamowski-R\"ossler system. Analytical results are 
confirmed convincingly in numerical simulations. The effect is quite general,
and occurs whenever the Floquet multipliers governing the stability of the 
limit cycle are complex, with the amplitude of the oscillations increasing 
as the instability boundary is approached.
\end{abstract}


\maketitle

\section{Introduction}

The subject of non-linear dynamics, with its wide range of tools and 
techniques, and its classification of the diverse types of behavior 
encountered, has in the last 20 or 30 years transformed our understanding of 
many models in the physical and biological sciences \cite{strogatz,guc83}. 
All these systems are subject to random perturbations, but the study of the 
effects that the noise has on a particular system, while still very significant
\cite{moss,risken,gardiner}, has not been nearly so extensive. Frequently 
the noise is added to the deterministic equations in a fairly ad hoc manner to 
obtain stochastic differential equations of the Langevin type. What is less 
often done is to start from a well-defined ``microscopic'' model defined by 
either a Markov chain or a master equation, and to treat the deterministic 
(macroscopic) limit of the model in a unified framework which also incorporates
the stochastic elements of the problem. In this paper we will develop such a 
treatment for a particular class of problems. Conventional tools used in 
deterministic nonlinear dynamics (for example, Frenet frames and Floquet 
analysis) will turn out to also have a role to play in the stochastic version 
of the model.

The particular class of problems we shall investigate will be those which have
a deterministic limit which, at large times, approaches a limit cycle in an 
oscillatory manner. That is, trajectories spiral into the limit cycle at 
large times. The motivation for studying such systems is the widespread 
interest that there has been in the analogous phenomena in systems which 
approach a fixed point in an oscillatory fashion. In this case, the effect 
of noise is, in many cases, to transform the oscillatory decay into a 
sustained oscillation about the fixed point. In this way the long-time 
behavior of the system is no longer a fixed point, but consists of 
stochastic oscillations which have a frequency which may be different to that 
which appears in the oscillatory decay in the deterministic version. The 
possibility of such an effect occurring has been discussed for some time
\cite{bartlett,nisbet}, but it is only in the last few years that a full
quantitative analysis has been given. The method has been applied to the study 
of stochastic oscillations in predator-prey systems \cite{alan1,pineda}, 
epidemiology \cite{alan2,STdG07,kuske}, chemical reactions in the 
cell \cite{alan3}, auto-catalytic reactions \cite{auto}, among others. One of 
the important aspects of these oscillations is that they have an amplitude of 
order $c/\sqrt{N}$ relative to the deterministic trajectory, where $N$ is the 
size of the system (the maximum number of individuals, molecules, etc, that 
may be put into the system) and $c$ is a constant which due to a resonance 
effect is usually quite large. This means that even for relatively large 
values of $N$, where the oscillations would be expected to be small and 
stochastic effects negligible, the relative amplitude can be of order one,
and so the fluctuations may dominate the dynamics. This effect is usually 
referred to as stochastic amplification, to avoid confusion with the very 
different effect of stochastic resonance \cite{ben81}.

The question that will interest us here is: does a similar phenomenon happen 
in other contexts, in particular when the stable state of a deterministic 
dynamical system is a limit cycle? Much less work has been done for this 
case as compared with the case of a fixed point, yet intuitively we would 
expect a similar effect to occur. In fact, the only previous studies we are 
aware of are by Wiesenfeld \cite{wie1,wie2}, who investigated the effects of 
noise on the stability of periodic attractors of various dynamical systems, 
such as the driven pendulum. He obtained analytical and numerical results on 
the power spectra of fluctuations about the limit cycles of such systems, but 
he adopted the approach that we mentioned above: by adding noise to the 
deterministic equations of motion. This is acceptable if the noise is external,
as he was envisaging, but if one wishes to understand the possible 
amplification of the underlying fluctuations due to intrinsic demographic 
stochasticity, then one needs to begin with the discrete dynamics, as we have 
already emphasized.

In a recent paper \cite{boland}, we have investigated a stochastic
model of the well-known Brusselator system, which has a limit cycle in the 
deterministic limit. However, in this model the approach to the limit cycle 
is not oscillatory. As in the case of fixed points and in the applications 
listed above, a precondition for finding sustained coherent oscillations is 
for the stable limit cycles to be approached in an oscillatory manner. For 
a fixed point this condition is that the eigenvalues of the stability matrix 
about the fixed point are complex. For a limit cycle, the analogous condition 
is that the Floquet multipliers for the equations describing the small 
deviations away from the periodic path are complex. Floquet multipliers are 
found to be real in the Brusselator system, as the number of degrees of 
freedom is not large enough to produce complex multipliers, and no coherent 
amplification phenomenon is observed. Part of the motivation for the work
described in \cite{boland} was to put the necessary tools in place, and to 
set the stage, for the investigation of model systems in which persistent 
oscillatory behavior about a limit cycle is to be expected. 

We begin with a two-dimensional system. If the system is autonomous one of the 
Floquet multipliers will have a value of unity, which, as we will see, implies 
that the remaining Floquet multiplier has to be real. This means that complex
Floquet multipliers can only be found in two-dimensional systems if they are 
non-autonomous. It is natural to achieve this by imposing an external periodic
driving, so as to induce a limit cycle as the steady state. In order to make 
contact with our previous paper \cite{boland} we here first study the 
Brusselator forced by an external periodic driving. As it turns, out this 
system does indeed have complex Floquet multipliers for a range of possible 
values for the parameters of the model. We then discuss an autonomous
system in three dimensions: the Willamowski-R\"ossler model, first introduced 
to describe chemical chaos. Finally, we consider a coupled set of two 
Brusselator systems as a four-dimensional illustration. Although we focus on 
these particular examples in the present paper, the formalism we will develop 
will hold in arbitrary dimensions and it will apply whether the system is 
autonomous or non-autonomous.

The outline of the paper is as follows. We begin in Sec.~\ref{sec:FB} with the 
forced Brusselator. By avoiding the technical complexities of working in
general dimensions, appealing to some of the results used in our previous 
paper on the unforced Brusselator \cite{boland}, and not having to use the 
Frenet frame in the analysis, we hope to provide a gentle introduction to 
the basic ideas. In Sec.~\ref{sec:WR} we extend the analysis to the 
Willamowski-R\"ossler model which introduces some extra features over and 
above those used in Sec.~\ref{sec:FB}, and in Sec.~\ref{coupledB} we carry out 
the full analysis for a system in an arbitrary number of dimensions and 
illustrate its use on the coupled Brusselator. We conclude in 
Sec.~\ref{conclude}. There are three mathematical appendices which cover the 
details of the formalism and some aspects of the calculations for the specific 
models considered in the earlier sections. 

\section{Forced Brusselator}
\label{sec:FB}
In this section we will study the Brusselator system, subject to an external 
periodic forcing. An analysis of the unforced model can be found 
in \cite{boland}, and much of the formalism remains unchanged. As it turns 
out, the introduction of the forcing actually simplifies some aspects of the 
dynamics as discussed below. While we re-iterate the main elements of the 
formalism and of the notation in the present paper, our previous 
work \cite{boland} may be consulted for specific details.

\subsection{Model definitions}
The Brusselator model is a relatively simple chemical system, composed of 
five different reactants ($A,B,C,X_1$ and $X_2$), and governed by the 
reactions \cite{haken,serra,tomita}
\BE\label{eq:r1}
A&\rightarrow&X_1+A,\\\label{eq:r2}
X_1&\rightarrow&\emptyset,\\\label{eq:r3}
X_1+B&\rightarrow&X_2+B,\\\label{eq:r4}
2X_1+X_2+C&\rightarrow&3X_1+C.
\EE

These reactions conserve the numbers of molecules of types $A, B$ and
$C$ in the system, while those of $X_1$ and $X_2$ are the dynamical
degrees of freedom. The role of the substances $A,B$ and $C$ is mainly to set 
the rates with which the first, third and fourth reaction occur, respectively.

The concentrations of the $A$ and $C$ molecules
will be held constant in time in all variations of the model that we
will consider, while the concentration of substance $B$ will be used
to apply an external driving force. The precise manner
in which this forcing is implemented will be detailed below. On the
deterministic level, the system is described by the following two coupled
ordinary differential equations \cite{haken,serra,tomita}
\BE\label{eq:bruss_general}
\dot x_1&=&1-x_1\left(1+b(t)-cx_1x_2\right),\nonumber\\
\dot x_2&=&x_1\left(b(t)-cx_1x_2\right),
\EE
where $x_1(t)$ and $x_2(t)$ describe the time-dependent concentrations
of substances $X_1$ and $X_2$ respectively, the constant $c$ the 
concentration of the $C$ molecules (the concentration of the $A$ molecules 
has been set equal to unity), and where $b(t)$ is the externally controlled 
concentration of $B$-molecules. The unforced Brusselator is recovered by 
setting $b(t)\equiv b_0$ independent of time. In this unforced case the system 
may exhibit both fixed points and limit cycles, depending on the choice of 
the coefficients $b_0$ and $c$ (see \cite{boland} and references therein for 
details), but no oscillatory approach to the limit cycles is possible as 
discussed below. For later convenience we rewrite Eqs.~(\ref{eq:bruss_general})
as $\dot{\bx} = \mathbf{A}(\bx,t)$ where 
\BE\label{eq:mean_spelled_out}
A_1(\bx,t)&=&1-x_1\left(1+b(t)-cx_1x_2\right),\nonumber \\
A_2(\bx,t)&=&x_1\left(b(t)-cx_1x_2\right).  
\EE 

To complete the definition of the model it remains to specify the functional 
form of the forcing. We will here use a deterministic, periodically varying 
forcing, $b(t)=b_0(1+\varepsilon\cos(\Omega t))$, in
Eqs.~(\ref{eq:bruss_general}). The non-negative control parameter $\varepsilon$
sets the amplitude of the external driving, and $\Omega$ is its angular 
frequency. We restrict ourselves to $\varepsilon<1$ so that the concentration 
of $B$-molecules remains non-negative. For $\varepsilon=0$ we recover the 
unforced Brusselator.

\subsection{Deterministic dynamics: Floquet analysis and stability of limit 
cycles}
\subsubsection{Floquet theory}
For periodic functions $b(t+T_\Omega)=b(t)$, and assuming $\varepsilon\neq 0$, 
any periodic solutions of Eqs.~(\ref{eq:bruss_general}) must have a time period
$T=nT_\Omega$, where $T_\Omega\equiv 2\pi/\Omega$ and $n$ is a positive 
integer. Numerical integration indeed shows that such cycles are found, though 
not for all choices of the model parameters. Furthermore, for the parameters 
that we have tested we only find limit cycles corresponding to $n=1$. The 
stability of these periodic solutions may then be analyzed within
the framework of Floquet theory \cite{grimshaw,strogatz}. Assuming model 
parameters are set such that a periodic solution, $\olx(t)$, of 
Eqs.~(\ref{eq:bruss_general}) exists, one considers a small perturbation, 
$\bxi(t)$, about this solution. To linear order one then has 
\be\label{eq:linear} 
\frac{\dd\bxi(t)}{\dd t}=K(t)\bxi(t), 
\ee 
where $K(t)$ denotes a $2\times 2$ matrix with entries
$K_{ij}(\olx(t),t)=\partial_j A_i(\olx(t),t)$, $i,j = 1,2$ and
$\partial_j$ denotes a derivative with respect to $\ovx_j$. The explicit form 
of $K(t)$ is given by Eq.~(\ref{eq:forcedK}) of Appendix \ref{app:a}, but 
given that $b(t)$ and $\olx(t)$ are of period $T_\Omega$ it follows that 
$K(t+T_\Omega)=K(t)$. Equation (\ref{eq:forcedK}) is identical to that for 
the unforced case \cite{boland}, except that here $b$ is replaced by a 
time-dependent function $b(t)$. 

In essence, Floquet theory states that, provided $X(t)$ is a fundamental 
matrix of (\ref{eq:linear}), then there exists a non-singular constant matrix 
$B$ such that 
\be\label{eq:FloquetB}
X(t+T_\Omega)=X(t)B,
\ee
for all $t$. While the Floquet matrix $B$ will, in general, depend on the 
choice of the particular fundamental matrix $X(t)$, its eigenvalues can be 
shown to be independent of this choice \cite{grimshaw}.  The eigenvalues of $B$
are usually referred to as the Floquet multipliers of the system 
(\ref{eq:linear}). For the forced Brusselator model there are two multipliers, 
and we denote them by $\rho_1$ and $\rho_2$ in the following. Since $B$ is 
real, if one of the multipliers is real, so is the other. This is the 
situation found in two-dimensional autonomous systems. Characteristic 
exponents $\mu_1$ and $\mu_2$ are then defined by $\rho_i=e^{\mu_i T_\Omega}$ 
for $i = 1,2$. 

We will now briefly discuss the properties of the resulting Floquet 
multipliers. In order to make contact with the unforced system it is
useful to distinguish between the cases $b_0<1+c$ and $b_0>1+c$, as the 
attractor of the unforced system is a stable fixed point in the former case, 
and a limit cycle in the latter \cite{boland}. 

\subsubsection{The case $b_0<1+c$}

\begin{figure}\center
\includegraphics[width=.8\columnwidth]{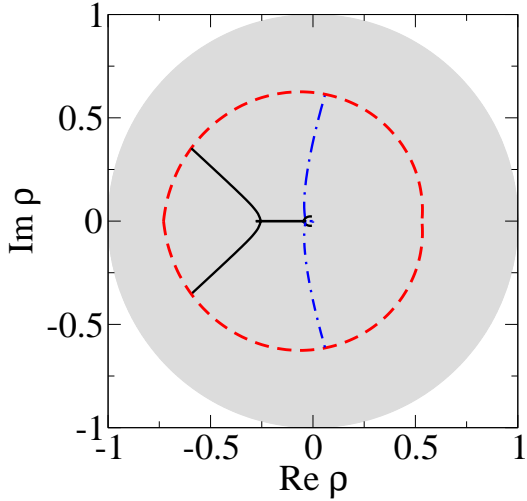}
\caption{\label{fig:epsilon1-8}
(Color online) An Argand diagram of the Floquet multipliers, $\rho_i$, for 
$b_0=1.8$ and $c=1$. The dashed red line indicates the position of the 
multipliers as $\varepsilon\to 0$, with 
$\tau\in [\pi/\omega^*,2\pi/\omega^*]$ (cf. Eq.~(\ref{eq:logspiral})). The 
blue dot-dashed line shows the location of the multipliers at $\Omega=1.3$, 
where different points on the line correspond to different values of 
$\varepsilon>0$. The solid line is for $\Omega=1.7$. The forcing amplitude is 
varied in the range $\varepsilon\in[10^{-2},1]$. The shaded area is the unit 
disk.}
\end{figure}

A trivial application of Floquet theory is to the unforced case
($\varepsilon\to0$), so that $b(t)\equiv b_0$. For $b_0<1+c$ the 
deterministic system is then known to approach a fixed point, see 
e.g. \cite{boland} for further details. Floquet theory remains formally 
applicable as the matrix $K(t)$ in Eq.~(\ref{eq:linear}) becomes 
time-independent at the fixed point; we will write $K(t)=K^*$. Indeed, in 
this case, formally the time period of the matrix $K$ can be set arbitrarily, 
as one has $K(t+\tau)=K(t)$ for all $\tau$ and $t$. Solutions to 
(\ref{eq:linear}) may be obtained directly by integration, and they can be 
written as $\bxi(t)=\exp\{K^*t\}\bxi_0$, where we have set the initial 
condition to be $\bxi(0)=\bxi_0$. Considering two solutions, generated from two
linearly independent initial conditions, we can construct a
fundamental matrix, $X(t)$. It then follows from the form of the
solutions to Eq.~(\ref{eq:linear}), and from Eq.~(\ref{eq:FloquetB}), that
the Floquet matrix $B$ depends on the choice of the period, $\tau$, as
\be
\label{eq:B_fixed} B(\tau)=\mathrm{e}^{K^* \tau}.  
\ee 
Denoting the eigenvalues of $K^*$ by $\lambda_i$, $i=1,2$, and those of 
$B(\tau)$ by $\rho_i(\tau)$, Eq.~(\ref{eq:B_fixed}) yields the relation
$\rho_i(\tau)=\exp\{\lambda_i \tau\}$. If the eigenvalues $\lambda_i$
are complex, then they are a complex conjugate pair, $\lambda_\pm$. Setting 
$c=1$ (which we do from now on), the eigenvalues of $K^*$ are given by 
$\lambda_\pm=(b_0/2)-1\pm\mathrm{i}\sqrt{b_0(4-b_0)}$, i.e. they are a pair 
of complex conjugates with non-zero imaginary part so long as $b_0<4$. The 
imaginary parts of $\lambda_{\pm}$ will be denoted by $\pm \omega^*$. We 
will refer to $\omega^*$ as the natural frequency of the unforced Brusselator.
When forcing is applied, then in the limit $\varepsilon\to 0$, the functions 
$\rho_i(\tau)=\rho_\pm(\tau)$ are logarithmic spirals in the complex plane, 
for $b_0<1+c$, i.e.~they are of the form
\be\label{eq:logspiral}
\rho_\pm(\tau)=\mathrm{e}^{[(b_0/2)-1]\tau}\left(\cos(\omega^* \tau)
\pm\mathrm{i}\sin(\omega^* \tau)\right).  
\ee 
Following Wiesenfeld \cite{wie2}, we illustrate the position of the Floquet 
multipliers in the complex plane on an Argand diagram, see 
Fig.~\ref{fig:epsilon1-8}. The dashed line here corresponds to 
Eq.~(\ref{eq:logspiral}) at $b_0=1.8$ for the range  
$\tau \in [(\pi/\omega^*),(2\pi/\omega^*)]$.


If the forcing amplitude $\varepsilon$ is small, but non-zero, the 
deterministic dynamics (\ref{eq:bruss_general}) no longer approaches a fixed 
point, but instead it is found to have a limit cycle. In this limit however, 
the deterministic trajectory is observed to remain close to the fixed point 
of the unforced case. The matrix $K(t)$ in Eq.~(\ref{eq:linear}) then
approaches $K^*$ as $\varepsilon \to 0$. It then follows that 
$\rho_i \to \exp\{\lambda_iT_\Omega\}$, so that the Floquet exponents 
$\mu_i \to \lambda_i$ as $\varepsilon \to 0$. We find from a numerical 
integration of the deterministic dynamics that increasing the level of the 
forcing amplitude tends to make the forced limit cycle more stable; that is, 
we find that the modulus of $\rho_1$ and $\rho_2$ decreases when $\varepsilon$ 
is increased, as shown in Fig.~\ref{fig:epsilon1-8}.

\begin{figure}
\includegraphics[trim=0mm 190mm 0mm 0mm,width=1\columnwidth]{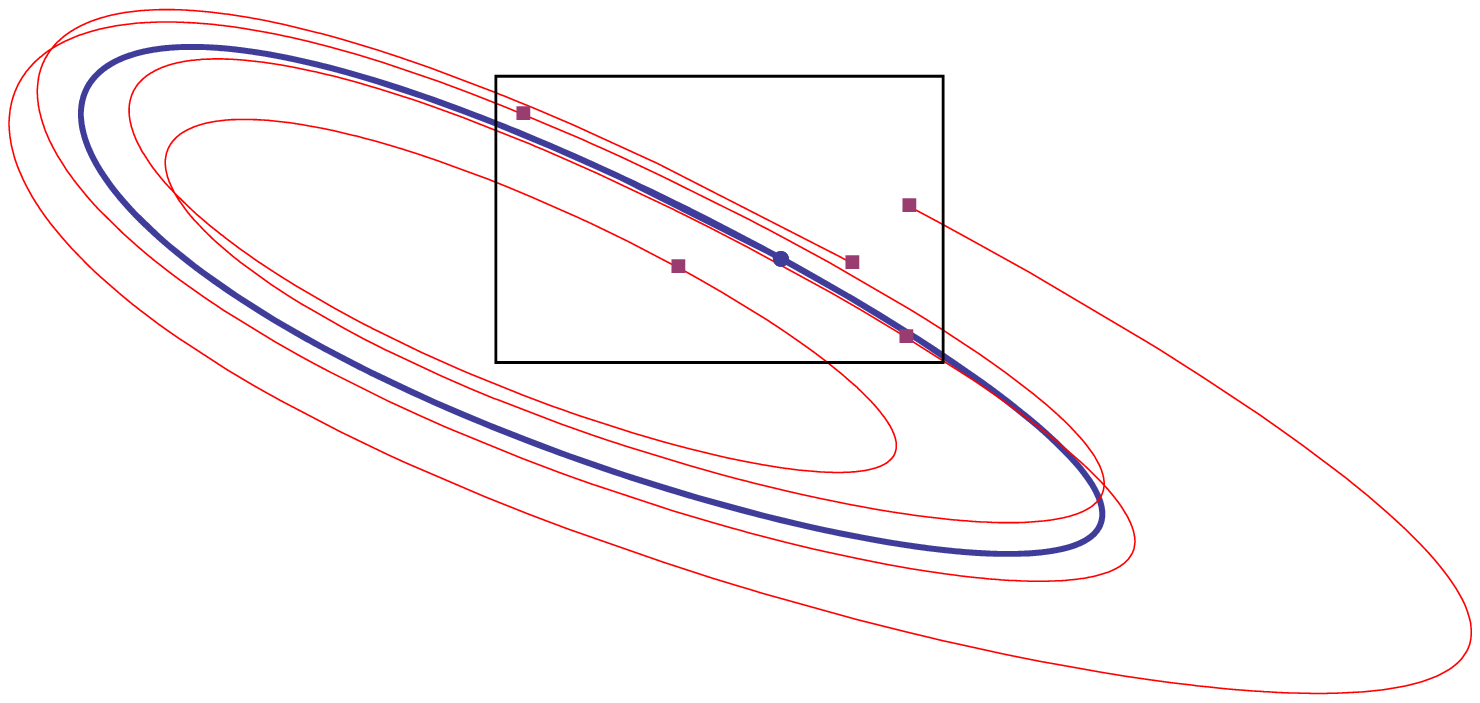}
\includegraphics[trim=0mm 160mm 0mm 0mm,width=.65\columnwidth]{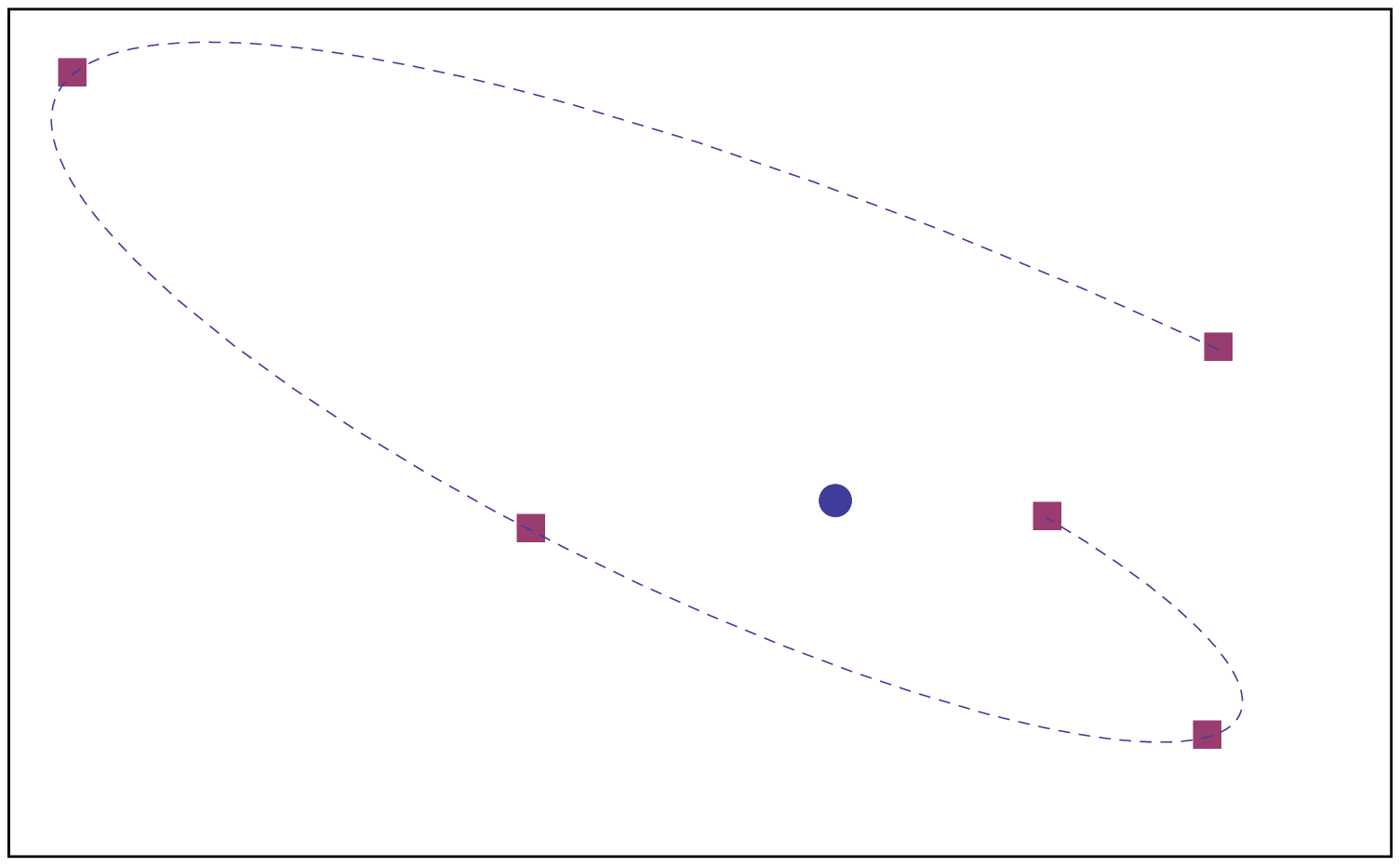}
\caption{\label{fig:schematic}
(Color online) The top panel is a schematic plot of a deterministic approach, 
shown as a thin gray (red) curve, to a limit cycle with complex Floquet 
multipliers; the cycle itself appearing as the closed dark curve (blue). The 
lower panel, showing a stroboscopic section, illustrates the spiraling return 
to the limit cycle with a frequency distinct from that of the limit cycle 
itself.}
\end{figure}

Let us end this subsection by returning to the interpretation of complex 
Floquet multipliers. According to Floquet theory, a solution to 
Eq.~(\ref{eq:linear}) may be written as a linear combination of solutions 
which have the property 
$\bxi_i(t+T_\Omega)=\rho_i\bxi_i(t)$ for $i = 1,2$. When the $\rho_i$ are 
complex conjugate pairs, this means that linear displacements from the 
periodic solution $\bolx(t)$ return to the limit cycle in elliptical spirals, 
in a way similar to the stable fixed point of the unforced case. We illustrate 
this typical behavior of complex Floquet multipliers in 
Fig.~\ref{fig:schematic}.


\subsubsection{The case $b_0>1+c$}
\begin{figure}\center
\vspace{3em}
\includegraphics[width=.8\columnwidth]{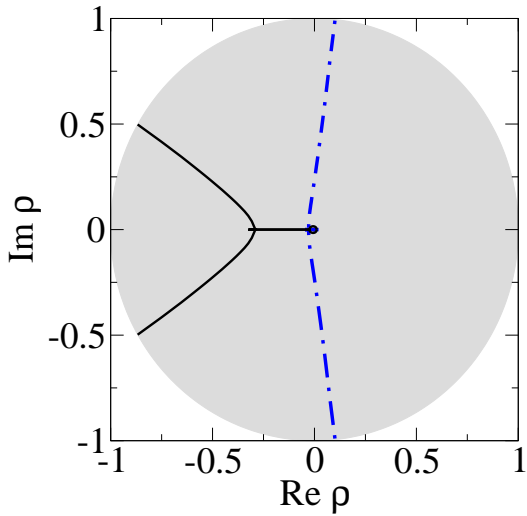}
\caption{\label{fig:epsilon2-2} (Color online) Location of the Floquet 
multipliers in the complex plane for $b_0=2.1$ and $c=1$. The dot-dashed blue 
line is for a forcing frequency of $\Omega=1.3$, the solid black line for 
$\Omega=1.7$. Periodic solutions are found above $\varepsilon_c=0.068$ and 
$\varepsilon_c=0.14$ respectively. Floquet multipliers are shown for 
$\varepsilon_c\leq\varepsilon\leq 1$ for both values of $\Omega$. The shaded 
area is the unit disk, Floquet multipliers approach the unit circle as 
$\varepsilon$ approaches $\varepsilon_c$ from above.}
\end{figure}

The case in which $b_0>1+c$ is slightly more complicated than the one in which 
the unforced deterministic system approaches a fixed point. For $b_0>1+c$ the 
unforced system has a stable limit cycle solution \cite{boland}; we will 
denote its angular frequency by $\omega_0$, where $\omega_0$ generally 
depends on $b_0$ and on $c$. One of the Floquet multipliers is 
equal to unity \cite{boland,tomita}, $\rho_1=1$, while the other one is found 
to be in the range $0<\rho_2<1$, consistent with a stable limit cycle 
attractor. We were not able to find any stable periodic solutions when 
integrating Eqs.~(\ref{eq:bruss_general}) at small, but non-zero, forcing
amplitudes $\varepsilon$ at generic forcing frequencies. At fixed values of 
$b_0$ and $c$, periodic solutions are however found for all $\Omega$ when the 
forcing amplitude exceeds a critical value, which we denote by 
$\varepsilon_c(\Omega)$, suppressing a potential dependence on $b_0$ and $c$. 
For $\varepsilon\geq \varepsilon_c(\Omega)$ these solutions are stable limit 
cycles, and the corresponding Floquet multipliers lie within the unit circle. 
Here we will exclusively focus on this regime. At
$\varepsilon=\varepsilon_c(\Omega)$ the multipliers have a modulus of
one, so that the cycle loses its stability, and as in the previous subsection, 
increasing the forcing amplitude reduces the moduli of $\rho_1$ and $\rho_2$, 
as shown in Fig.~\ref{fig:epsilon2-2}.
For our purposes it is sufficient to go on to study the case where the 
Floquet multipliers remain inside the unit circle, and to analyze the power 
spectra of stochastic fluctuations about the limit cycle in this regime.

\subsection{Stochastic dynamics and system-size expansion}
\subsubsection{Specification of the Model}
We now turn to a discussion of the stochastic microscopic Brusselator system, 
as defined by the reactions (\ref{eq:r1})-(\ref{eq:r4}). Labeling the 
reactions by $\nu = 1,\ldots,4$, we denote the rates with which each of the
reactions occur by $T_\nu(\bn,t)$. These rates depend on the state of the 
system $\bn=(n_1,n_2)$, where $n_i$ is the number of molecules of species 
$X_i$, and for the forced system have an additional explicit dependence on 
time. For the Brusselator system
$T_1=N$, $T_2(\bn)=n_1$, $T_3(\bn,t)=b_0(1+\epsilon\cos(\Omega t))n_1$ and 
$T_4(\bn)=cN^{-2}n^2_1n_2$. The combinatorial factors are as in the unforced
case \cite{boland}. The time-dependent expression for $T_3$ reflects the 
periodic forcing, implemented through an externally-controlled variation of 
the number of $B$-molecules in the system. We also define the vectors 
$\bv_\nu$, $\nu=1,\dots,4$, each capturing the effects of a single 
occurrence of a reaction of type $\nu$ on the numbers of $X_1$ and $X_2$ 
molecules in the system. For the Brusselator 
$\bv_1=(1,0)$, $\bv_2=(-1,0)$, $\bv_3=(-1,1)$ and $\bv_4=(1,-1)$ \cite{boland}.

\subsubsection{Analytical description and system-size expansion}
The time evolution of the probability, $P_\bn(t)$, of finding the system in 
state $\bn$ at time $t$ is then governed by the master equation
\be\label{eq:chemical_master}
\frac{\dd P_\bn(t)}{\dd t}=\sum_{\nu=1}^4\left[T_{\nu}(\bn-\bv_{\nu},t)
P_{\bn-\bv_\nu}(t)-T_{\nu}(\bn,t)P_\bn(t)\right].
\ee
Solving the master equation analytically is generally not feasible, but an 
effective description in terms of a Langevin equation, valid at large, but 
finite, system size can be obtained by means of a van Kampen expansion in the 
inverse system size \cite{vankampen}. 

This procedure is well-established and has been applied to a number of
microscopic interacting particle systems, so that we do not describe the
mathematical details here, but instead refer to \cite{vankampen,alan3}. The 
main idea is to expand realizations $\bn(t)$ of the microscopic dynamics about 
a deterministic trajectory, $\olx(t)$, 
\be\label{eq:vk}
\frac{\bn(t)}{N}=\olx(t)+\frac{1}{\sqrt{N}}\bxi(t),
\ee
and to derive an equation of motion for the fluctuations, $\bxi(t)$, from an 
expansion of the master equation (\ref{eq:chemical_master}) in powers of 
$N^{-1/2}$. To lowest order one finds that self-consistency requires 
$\dot \olx=\mathbf{A}(\olx,t)$, where
$\mathbf{A}(\bx,t)=(A_1(\bx,t),A_2(\bx,t))$ is given 
by the expressions in Eq.~(\ref{eq:mean_spelled_out}), recovering the 
deterministic dynamics of Eqs.~(\ref{eq:bruss_general}). These equations 
may also be derived by defining
\be\label{eq:ensemble}
\avg{\bn(t)}=\sum_{\bn}\bn P_\bn(t),
\ee
and noting that
\be\label{eq:mean-field}
\frac{\dd \avg{\bn(t)}}{\dd t}=\sum_{\nu=1}^4\bv_{\nu}T_{\nu}
\big(\avg{\bn(t)},t\big),
\ee
where we have used a deterministic approximation to write 
$\avg{T_\nu(\bn,t)}=T_\nu(\avg{\bn(t)},t)$. 
Equations~(\ref{eq:mean_spelled_out}) are then recovered by setting 
$\bx(t)=\avg{\bn(t)}/N$. At next-to-leading order the van Kampen expansion 
gives a linear Langevin equation for the fluctuations, $\bxi(t)$, about the 
deterministic trajectory which has the general form \cite{vankampen,alan3}
\be\label{eq:lang_forcedbruss}
\frac{\dd \bxi(t)}{\dd t}=K(t)\bxi(t)+\beff(t),
\ee
where, for the forced Brusselator, the matrix $K(t)$ is defined in 
Eq.~(\ref{eq:forcedK}). The term $\beff(t)$ on the right-hand side represents 
a Gaussian noise of zero mean and with correlator
\be\label{eq:matrixd}
\avg{f_i(t)f_j(t')}=2D_{ij}(t)\delta(t-t').
\ee
The matrix $D(t)$ may be straightforwardly calculated from the van Kampen 
expansion \cite{vankampen,alan3}. The explicit form for the forced 
Brusselator is given by Eqs.~(\ref{eq:forcedD}) and (\ref{apeq:D}) in 
Appendix \ref{app:a}.

Equation (\ref{eq:lang_forcedbruss}) is a linear Langevin equation, and
analytical progress is therefore possible. Of particular interest to us here 
are the correlation functions and power spectra of the fluctuations 
$\bxi(t)$. The time-averaged elements of the covariance matrix 
$C_{ij}(t,t')=\avg{\xi_i(t)\xi_j(t')}$ are defined as
\be\label{eq:forced_auto}
C_{ij}(\tau)= \frac{1}{T_{\Omega}} \int^{T_{\Omega}}_{0} dt\,
\left\langle \xi_i(t) \xi_j(t+\tau) \right\rangle.
\ee
We will in the following mostly focus on the diagonal elements $C_{ii}(\tau)$. 
Even though Eq.~(\ref{eq:lang_forcedbruss}) is linear, the analytical 
computation of $C_{ii}(\tau)$ requires several intermediate steps, and final 
expressions need to be evaluated numerically. The details are left until the 
general theory, applicable to systems in an arbitrary number of dimensions, 
is explained in Sec.~\ref{coupledB}.

\subsubsection{Comparison with simulations}
\begin{figure}\center
\includegraphics[width=0.9\columnwidth]{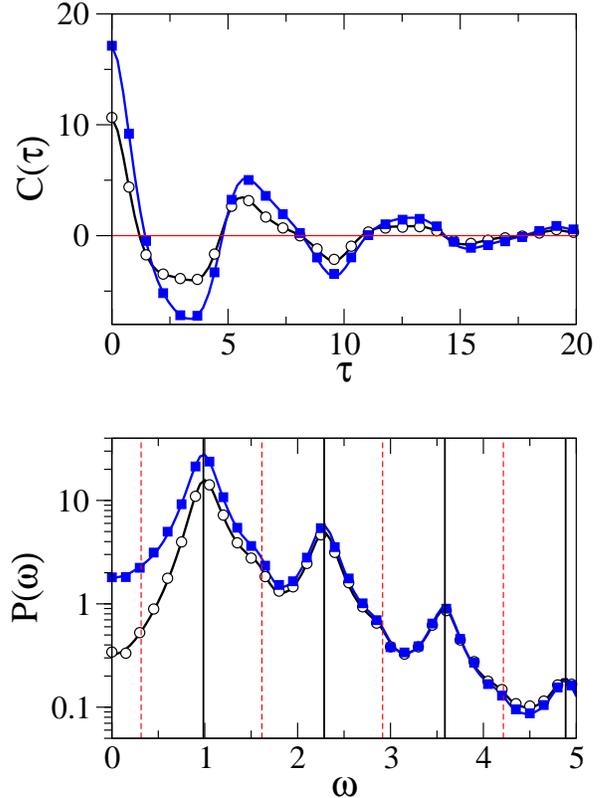}
\vspace{1em}
\caption{\label{fig:forcedbruss} (Color online) The autocorrelation
(top panel) and the power spectra (bottom panel) of stochastic fluctuations 
about the deterministic trajectory of the forced Brusselator. Simulation 
results for fluctuations $\xi_1$ of the number of $X_1$-molecules are shown as 
open circles, while those for $\xi_2$ are indicated by full squares. The solid 
lines are the predictions of the theory, and are seen to match the simulations 
perfectly. Model parameters are $b_0=2.1, c=1, \Omega=1.3$ and 
$\varepsilon=0.14$. The corresponding non-trivial Floquet multipliers are 
$\rho=0.023\pm0.46i$, so that $|\rho|=0.46$. The imaginary part of the Floquet 
exponent is $\mbox{Im}~\mu=0.32$. The system size in simulations is 
$N=2 \times 10^5$ and averages over $5000$ independent runs are taken. 
Vertical lines in the lower panel mark the frequencies of $n\Omega-{\rm Im}\mu$
(solid lines) and $n\Omega+{\rm Im}\mu$ (dashed), where $n$ is a positive 
integer.}
\end{figure}

In Fig.~\ref{fig:forcedbruss} we compare results from the analytical 
calculation just described, with measurements obtained from simulations of 
the microscopic dynamics. Simulations are carried out using the 
Gillespie algorithm \cite{gil77}, suitably modified to account for the 
explicit time-dependence of the reaction rates induced by the external forcing 
\cite{and07}. Measurements in simulations are taken after a suitable 
equilibration period in order to minimize the effects of transients. 
Fig.~\ref{fig:forcedbruss} shows results from the theory (lines) and from 
simulations (markers), and as seen in the figure, the agreement between them 
is excellent, both for the time-averaged autocorrelation functions 
$C_{ii}(\tau)$ and the corresponding power spectra. The latter are obtained 
as the Fourier transforms of the correlation functions: 
\be\label{eq:Fourier}
P_i(\omega)=\int~ d\tau e^{i\omega\tau}C_{ii}(\tau).
\ee


\begin{figure}\center
\includegraphics[width=.95\columnwidth]{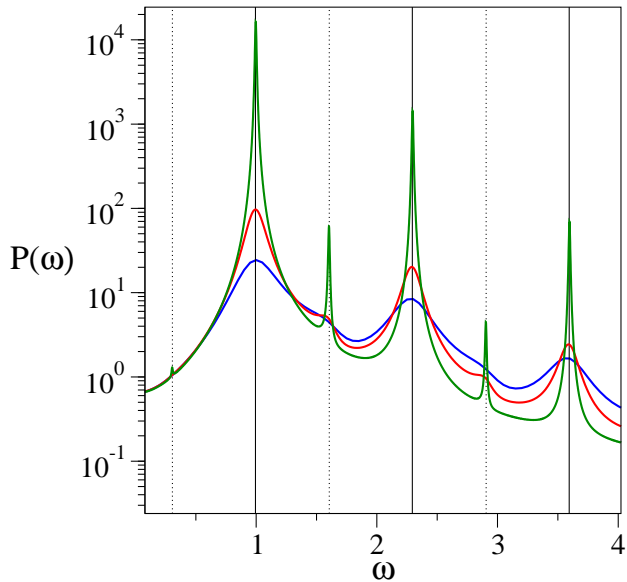}
\vspace{1em}
\caption{\label{fig:spectrum_extreme} (Color online) Power spectrum of 
stochastic fluctuations, $\xi_1$, in the forced Brusselator system as obtained 
from the analytical calculations. Model parameters are $b_0=2.1$, $c=1$, 
$\Omega=1.3$, resulting in $\varepsilon_c=0.068$. The different curves 
correspond to forcing amplitude $\varepsilon=0.07, 0.11, 0.15$, from top to 
bottom, at the peaks. The corresponding Floquet multipliers have modulus 
$0.97, 0.66$ and $0.41$ respectively. Vertical lines are given at frequencies 
of $n\Omega-{\rm Im}\mu$ (solid lines) and $n\Omega+{\rm Im}\mu$ (dotted),  
where $n$ is a positive integer and where $\mbox{Im}\mu\approx0.31$ for all 
three cases.}
\end{figure}

In the numerical simulations we first measure $C_{ii}(t,t')$, and then perform 
a time-average to obtain $C_{ii}(\tau)$. Subsequently a discrete Fourier 
transform is taken, to obtain $P_i(\omega)$. From a practical point of view,
$C_{ii}(\tau)$ is found only for $\tau \geq 0$, and then the even nature of
the function (discussed later) invoked. Wiesenfeld \cite{wie1} suggested 
peaks would be expected to be seen at frequencies $n\Omega \pm {\rm Im}\mu$, 
where $n$ is a positive integer and ${\rm Im}\mu$ is $| {\rm Im} \mu_{1,2}|$,
where $\mu_{1,2}$ are the two Floquet exponents. However, our results indicate 
that the presence or otherwise of such peaks depends strongly on the choice of 
model parameters, and in particular on the position of the Floquet multipliers 
in the complex plane. For the case shown in Fig.~\ref{fig:forcedbruss}, for 
example, $\rho_{1,2}=0.023\pm0.46i$ and marked peaks are found at 
$n\Omega - {\rm Im}\mu$, but not at $n\Omega + {\rm Im}\mu$. A second example 
is shown in Fig.~\ref{fig:spectrum_extreme}, where we show data for a number 
of model parameters, resulting in Floquet multipliers much closer to the unit 
circle than for the example shown in Fig.~\ref{fig:forcedbruss}. Peaks are now 
found at all $n\Omega \pm {\rm Im}\mu$, with the peaks becoming more 
pronounced as the Floquet multipliers approach the unit circle (from within). 
In the limit $|\rho_{1,2}|\to 1$, the relaxation of autocorrelation functions 
becomes very slow and so larger values of $\tau$ need to be taken into account
when performing the Fourier transform. This makes both the analytical 
expressions and the Gillespie simulations more computationally expensive and, 
for the parameters illustrated in Fig.~\ref{fig:spectrum_extreme}, Gillespie 
simulation is not feasible.


\section{Willamowski-R\"ossler system}
\label{sec:WR}
\subsection{Microscopic model}
We have seen that forcing the two-dimensional Brusselator opens up the
possibility of complex Floquet multipliers. This was not possible in the 
unforced case since there the deterministic dynamics is autonomous, leading 
directly to a Floquet multiplier of unity. Therefore, in order to see 
the effects of complex Floquet multipliers in an autonomous system, the 
simplest case has three dimensions. One such system is the 
Willamowski-R\"ossler model and we shall study the particular form given in
\cite{geysermans1,geysermans2}. The model may be written as a chemical 
reaction system, involving three species $X_1$, $X_2$ and $X_3$, defined by
\BE\label{eq:logistic1}
X_1&\reversible{b_1}{d_1}&2X_1,\\\label{eq:logistic2}
X_2&\reversible{b_2}{d_2}&2X_2,\\\label{eq:annihilation}
X_1+X_2&\xrightarrow{1}&\emptyset,\\\label{eq:infection}
X_1+X_3&\xrightarrow{1}{}&2X_3,\\\label{eq:death}
X_3&\xrightarrow{d_3}&\emptyset.
\EE
The parameters above and below the arrows indicate the relative rates with 
which each of the reactions occur. Absorbing potential combinatorial factors 
in the definition of the model parameters $b_1,b_2,d_1$, and $d_2$, the four 
reactions given in Eqs.~(\ref{eq:logistic1}) and (\ref{eq:logistic2}) occur 
with rates $T_1(\bn)=b_1n_1$, $T_2(\bn)=d_1n_1^2N^{-1}$, $T_3(\bn)=b_2n_2$ and
$T_4(\bn)=d_2n_2^2N^{-1}$. In isolation, these four reactions, 
(\ref{eq:logistic1}) and (\ref{eq:logistic2}), ensure that the average 
numbers of species $X_1$ and $X_2$, are of the order of $N$, so that 
$N$ is again a measure of the system size. We will take the annihilation 
process (\ref{eq:annihilation}) to occur with rate $T_5(\bn)=n_1n_2N^{-1}$;
the prefactor in the rate of the reaction is taken to equal unity in order to 
agree with \cite{geysermans1,geysermans2}. The mathematically interesting 
limit is that in which the number of $X_3$ particles, $n_3$, is of order $N$ 
as well. This is the case when the remaining reaction rates are scaled 
suitably with $N$. Specifically we will assume that (\ref{eq:infection}) 
occurs at rate $T_6(\bn)=n_1n_3N^{-1}$ and (\ref{eq:death}) at rate 
$T_7(\bn)=d_3n_3$. The vectors, $\bv_\nu$, that correspond to the reactions 
$\nu = 1,\ldots,7$ are given by
\BE\nonumber
\bv_1=(1,0,0),&&\bv_2=(-1,0,0),\\\nonumber
\bv_3=(0,1,0),&&\bv_4=(0,-1,0),\\\nonumber
\bv_5=(-1,-1,0,)&&\bv_6=(-1,0,1),\\
\bv_7=(0,0,-1).&&
\EE

\subsection{Deterministic Dynamics and Frenet Frame}
As in the case of the forced Brusselator, we may now find the equations of
the corresponding deterministic dynamics using Eq.~(\ref{eq:mean-field}). For
the Willamowski-R\"ossler model these are
\BE\label{eq:WR}
\dot{x}_1=A_1(\bx)&=&x_1(b_1 - d_1x_1 - x_2 - x_3),\\
\dot{x}_2=A_3(\bx)&=&x_2(b_2 - d_2x_2 - x_1),\\
\dot{x}_3=A_3(\bx)&=&x_3(x_1 - d_3).
\EE
There are a total of six fixed points of this system, but only 
one at which all concentrations are non-zero. This fixed point is given by
\be
\bx^*=\left(d_3,\frac{b_2-d_3}{d_2},b_1-d_1d_3-\frac{b_2-d_3}{d_2}\right).
\ee
The stability matrix, $K_{ij}(\bx)=\partial_jA_i(\bx)$, at this fixed point may
be found from Eqs.~(\ref{eq:will_K1}) and (\ref{eq:will_K2}) in 
Appendix \ref{app:a}, by setting  $\olx(t) = \bx^{*}$. If the above non-trivial
fixed point is unstable then limit cycle solutions of the deterministic 
equations may exist. Such solutions have, for example, been reported 
in \cite{geysermans1,geysermans2}, and we will focus on this limit-cycle 
regime in this section.

Following the notation of the previous sections we will denote the 
deterministic limit cycle trajectory by $\olx(t)$, and we will write 
$\bxi(t)$ for the fluctuations about it, again as before. Much of the 
formalism we require has either been discussed in Sec.~\ref{sec:FB}, or 
in \cite{boland}. In particular one has an equation of the 
form (\ref{eq:linear}) within a linear stability analysis of the limit cycle,
and in the absense of noise. A direct consequence of the system being 
autonomous is that the velocity vector, $\dot{\olx}(t)$, is itself a solution 
to Eq.~(\ref{eq:linear}). Since the velocity is periodic, one of the Floquet 
multipliers is equal to unity, as it is generally the case for limit cycles 
of autonomous systems. The dynamics is marginally stable in the direction of 
the velocity, so that longitudinal fluctuations behave diffusively and may 
grow without bound in the long run \cite{boland, tomita}. We will focus our 
interest instead on the fluctuations in the transverse directions, since it 
is these that have the oscillatory behavior of interest to us. For stable 
limit cycles and in the absence of persistent noise, these transverse 
fluctuations decay in a manner characterized by the remaining Floquet 
multipliers. If the latter are complex, and if the system is subject to 
intrinsic noise, as induced by the underlying microscopic dynamics at finite 
system sizes, we expect these fluctuations to be enhanced into quasi-cycles 
about the limit cycle.

In order to separate longitudinal from transverse modes we need to introduce 
a suitable frame of reference. Such co-ordinates are provided by the Frenet 
frame \cite{gibson}, which may be constructed by applying the Gram-Schmidt 
orthogonalization procedure to the first three time derivatives of the limit 
cycle solution $\olx(t)$. Specifically, the co-moving basis vectors 
$\hat\bolde_i(t),\,i=1,2,3$ of the Frenet frame are constructed sequentially, 
as discussed further in Appendix \ref{app:b}. The fluctuations are governed 
by a Langevin equation of the form (\ref{eq:lang_forcedbruss}). In order to 
isolate the transverse fluctuations, we rotate the Langevin equation into 
the Frenet frame. After the rotation, defined by a matrix, $J(t)$, the 
Langevin equation takes the form 
\be\label{eq:rot_lang}
\dot{\bq}(t)=K^\mathrm{tot}(t)\bq(t)+\bg(t),
\ee
where we follow our earlier paper \cite{boland} and write $\bq(t)=J(t)\bxi(t)$
for the fluctuations in the Frenet frame. The matrix is periodic and given by
$K^\mathrm{tot}(t)=J(t)K(t)J^{-1}(t)+\dot{J}(t)J^{-1}(t)$ (see 
Appendix \ref{app:b}) and $\bg(t)=J(t)\beff(t)$ is the rotated noise term. It
follows from Eq.~(\ref{eq:matrixd}) that the components of $\bg(t)$ are each 
Gaussian white noise variables with zero mean and correlators
\be
\label{eq:matrixg}
\avg{g_i(t)g_j(t')}=2 G_{ij}(t)\delta(t-t'),
\ee
where $G(t)=J(t)D(t)J^{-1}(t)$. 

For autonomous systems, it is shown in Appendix \ref{app:b} that the 
existence of a longitudinal direction as described above, implies that the 
elements of the first column of the matrix $K^\mathrm{tot}(t)$ vanish, except 
for the entry in the first row. A consequence of this is that the transverse 
dynamics may be effectively considered independently of the dynamics in the
longitudinal direction. For the Willamowski-R\"ossler limit cycle this yields 
a pair of coupled linear Langevin equations in the two transverse directions,
with exactly the same mathematical form as those of the forced Brusselator 
model. Hence the same techniques as before may be applied to produce analytical
curves for the autocorrelations and power spectra in the two transverse 
directions. Note that in our previous work \cite{boland}, we were able to 
simplify the rotated Langevin equations further by a rescaling of the
coordinates in the Frenet frame. We do not apply this additional transformation
here, since for our purposes it is not essential.

\subsection{Stochastic Simulation and Results}
\label{subsec:stoch}
The Gillespie algorithm can again be used to generate realizations of the 
microscopic dynamics defined by Eqs.~(\ref{eq:logistic1})-(\ref{eq:death}). 
Since one Floquet multiplier in the Willamowski-R\"ossler system is equal to  
unity, there is a diffusive mode in the longitudinal direction. This means that
the time-evolution $(n_1(t)/N,n_2(t)/N,n_3(t)/N)$ of any single realization of 
this stochastic process may not remain close to the deterministic trajectory 
$\olx(t)$, but instead $\avg{|\bn(t)/N-\olx(t)|^2}\sim t$, where $|\cdot|$ 
stands for the Euclidean norm. This complication is not present in the driven 
Brusselator discussed in Sec.~\ref{sec:FB}, since in that case no such 
longitudinal diffusive mode exists.

\begin{figure}\center
\includegraphics[width=.95\columnwidth]{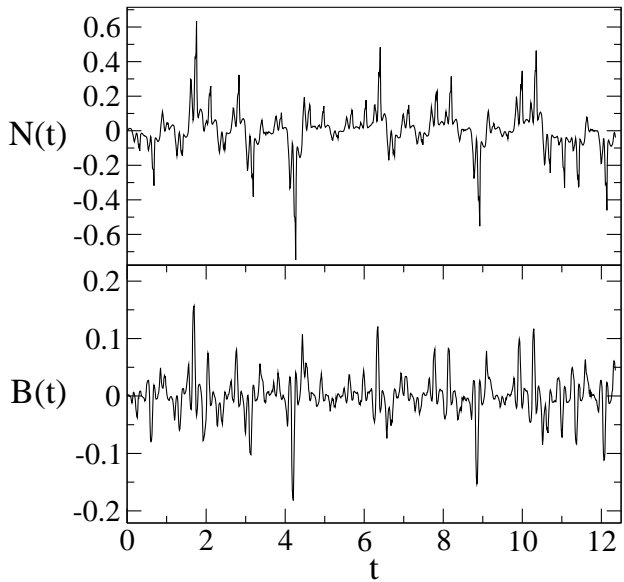}
\vspace{1em}
\caption{\label{fig:willexample} The fluctuations in the directions transverse 
to the limit cycle trajectory in the Willamowski-R\"ossler model. Data for the 
normal component, $N(t)$, and the binormal direction, $B(t)$, are shown for a 
single realization of the stochastic simulation. Model parameters are
$b_1=80$, $b_2=20$, $d_1=0.16$, $d_2=0.13$, and $d_3=16$. }
\end{figure}

This issue can however be dealt with as discussed in \cite{boland}. The 
procedure of extracting the deviation from the limit cycle is as follows: 
for every given data point $\bn(t)/N$ generated by the Gillespie 
algorithm one identifies the point $\olx(\bn(t))$ on the limit cycle 
trajectory which is geometrically closest to $\bn(t)/N$, and then uses
$\bkappa(t)=\bn(t)/N-\bolx(\bn(t))$ as the displacement vector. As described 
in \cite{boland} the longitudinal component of $\bkappa(t)$ vanishes, i.e. 
one has $\dot\olx.\bkappa=0$, while the remaining components define a 
stochastic process in the co-moving transverse plane, and as seen in 
\cite{boland} the magnitude of $\bkappa$ remains of order $N^{-1/2}$. This 
procedure allows one to effectively decouple the diffusive longitudinal mode 
from the transverse ones, and we will focus on the transverse components in 
the following, in order to characterize stochastic oscillations about the 
deterministic limit cycle. These components are then expressed in the Frenet 
co-ordinates, defined at $\bolx(\bn(t))$. As an illustration, trajectories of 
the transverse components obtained from a single realization of the microscopic
dynamics are shown in Fig.~\ref{fig:willexample} for a fixed set of model 
parameters. In this figure, $N(t)$ denotes the normal component, 
$N(t)=\bkappa(t).\hat{\bolde}_2(t)$, and $B(t)$ denotes the deviation from the 
limit cycle in the binormal direction, $B(t)=\bkappa(t).\hat{\bolde}_3(t)$. 
Recall here that $\hat{\bolde}_2$ and $\hat{\bolde}_3$ define a co-moving 
frame, i.e. that they carry a time-dependence as well. 


\begin{figure}\center
\includegraphics[width=.95\columnwidth]{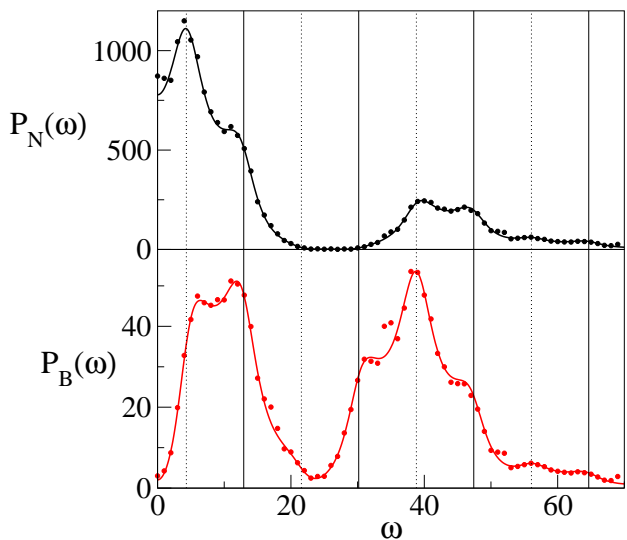}
\vspace{1em}
\caption{\label{fig:comp_spec} (Color online) Comparison of the theoretical 
and simulated estimates for the power spectra of transverse fluctuations in the
Willamowski-R\"ossler system. The top panel shows the normal fluctuations (in 
the direction $\hat{\bolde}_2)$ while the bottom panel compares those in the 
binormal direction, $\hat{\bolde}_3$. Model parameters are again set to 
$b_1=80$, $b_2=20$, $d_1=0.16$, $d_2=0.13$, and $d_3=16$. Vertical lines are 
given at frequencies $n\omega_0+\mathrm{Im}\,\mu$ (dotted) and 
$n\omega_0-\mathrm{Im}\,\mu$ (solid), with $n$ a positive integer. The 
numerical value of $\omega_0$ is $17.25$, and the non-trivial Floquet 
multipliers are $\rho=-0.002\pm0.303$ (resulting in $|\rho|=0.30$ and 
$\mbox{Im}~\mu=4.33$).}
\end{figure}


In Fig.~\ref{fig:comp_spec}, we show the resulting power spectra, and find very
good agreement between simulation and theory for both the normal and binormal 
directions. There is a slight systematic deviation of data points from the 
theory, which occurs at integer multiples of the limit cycle frequency. We 
attribute these to remnants of the deterministic dynamics. The data shown in 
Fig.~\ref{fig:comp_spec} was taken at model parameters resulting in complex 
Floquet multipliers with a modulus of approximately $0.3$, and peaks are found
in the power spectra close to frequencies $n\omega_0\pm\mathrm{Im}\,\mu$, where
$\omega_0$ is the angular frequency of the limit cycle. However, we also note 
that peaks are not observed at all frequencies $n\omega_0\pm\mbox{Im}\mu$, 
especially in the spectrum of normal fluctuations. As for our findings in the 
driven Brusselator, this may be due to the fact that the Floquet multipliers 
in the example shown in Fig.~\ref{fig:comp_spec} are relatively distant from 
the unit circle in the complex plane. Again based on our observations in the 
driven Brusselator one may expect additional peaks at frequencies 
$n\omega_0\pm\mbox{Im}\mu$ to emerge as the Floquet multipliers move closer 
to the unit circle. Despite an extensive search we have however not been able 
to find a set of model parameters which would result in Floquet multipliers 
with modulus close to unity, so that we are not able to give any further 
confirmation of this expectation here. We conclude this section by re-iterating
our main result, the near perfect agreement of the analytically obtained 
power spectra with simulations as shown in Fig.~\ref{fig:comp_spec}.  

\section{Generalization to higher dimensions and the coupled Brusselator}
\label{coupledB}
\subsection{General theory}
It is expected that in the study of any real system, for example in
biochemistry or in ecology, the number of distinct species, $S$, would be
significantly larger than two or three. It is also possible that solutions to 
the $S$-dimensional deterministic equations in such a model may be periodic 
orbits, $\bolx(t)$. Hence, in this section we demonstrate the natural extension
of the analysis in the previous sections to models of arbitrary dimension. 
Whether or not the system is autonomous, we begin with the van Kampen 
system-size expansion which yields a set of $S$ coupled and linear Langevin 
equations for stochastic fluctuations, $\bxi(t)$. We simply note that 
their form naturally extends to arbitrary dimension and is unchanged from
(\ref{eq:lang_forcedbruss}), where the $S\times S$ matrix $K(t)$ for the 
drift is given by 
$K_{ij}(t)=K_{ij}(\bolx(t))=\partial A_i(\bolx(t))/\partial \ovx_j$, and the 
symmetric $S\times S$ matrix for diffusion, $D(t)=D(\bolx(t))$, is calculated 
from the system-size expansion.

The subsequent steps of the analysis then depend on whether the system 
under consideration is autonomous or not. For non-autonomous systems no 
rotation is required, and one proceeds directly with the Langevin equation 
in Cartesian co-ordinates in $S$ dimensions. If the system is autonomous, 
as in the case of the Willamowski-R\"ossler model, one first needs to 
rotate into the $S$-dimensional Frenet frame, and then to separate off the 
longitudinal component, resulting in a Langevin equation in $S-1$ 
dimensions for the transverse components. The Frenet frame is defined in 
$S$ dimensions in Appendix \ref{app:b}. This then specifies the rotation
matrix $J^\mathrm{T}=(\hat{\bolde}_1,\ldots,\hat{\bolde}_S)$, which is
evaluated on the limit cycle so that $J(t)=J(\bolx(t))$. The formalism is
a straightforward generalization of that described in Sec.~\ref{sec:WR} for
the Willamowski-R\"ossler model, except that there are now $(S-1)$ transverse 
directions, rather than just two.

Thus, for both autonomous and non-autonomous systems one eventually ends up 
with a Langevin equation in $d$ dimensions, where $d=S-1$ for the autonomous 
case, and $d=S$ for non-autonomous systems, such as the driven Brusselator. 
The further steps of the calculation can hence be discussed simultaneously 
for the autonomous and non-autonomous cases. As described in more detail in 
Appendix \ref{app:c}, the solution of this Langevin equation can be expressed 
in terms of any fundamental matrix $X(t)$ of the corresponding homogeneous 
equation. Since the drift matrix, $K(t)$ (denoted by $\widetilde{K}(t)$ in the 
autonomous case) is periodic, Floquet theory \cite{grimshaw} asserts that a 
canonical fundamental matrix may be written in the form $X(t)=P(t)Y(t)$, where 
$P(t)$ and $Y(t)$ are $d\times d$ matrices. The matrix $P(t)$ is periodic with 
the same period as the drift matrix while the matrix $Y(t)$ is given by 
$Y(t)=e^{\mathrm{diag}\{\mu_1,\ldots,\mu_d\}t}$, where the 
$\mu_i$, $i=1,\ldots,d$ are the Floquet exponents of the $d\times d$ 
homogeneous system.

The periodic matrix, $P(t)$, acting on the left is, in effect, a transformation
matrix from the Floquet solutions to the coordinates of the Langevin equation, 
while its inverse makes the reverse transformation. The matrix $Y(t)$ is 
a diagonal exponential matrix with entries $e^{\mu_i t}$, with 
${\rm Re}\mu_i < 0$, for all $i$, for a stable limit cycle. It therefore acts 
on different Floquet solutions in different ways, reducing the value of some 
more quickly than others. The general solution of the Langevin 
equation (\ref{eq:transverse_Lang}) which we wish to analyze, can be given in 
terms of the matrices $P(t)$ and $Y(t)$ and is given explicitly by 
Eq.~(\ref{eq:stationary_solution}) in Appendix \ref{app:c}. 

Given the symmetric and periodic noise matrix, $D(t)$ in the non-autonomous 
case---which we generally denote by $\widetilde{G}(t)$ using the notation of
the autonomous case---we may calculate the autocorrelation function in closed 
form. In the basis corresponding to the Floquet solutions 
$\widetilde{G}(t)$ becomes the symmetric and periodic matrix
$\Gamma(t)=P^{-1}(t)\widetilde{G}(t)\big(P^{-1}\big)^\mathrm{T}$. These noise
contributions are then integrated over one time period of the deterministic
limit cycle, $T$, but weighted by decaying exponentials from the $Y(t)$
matrix, to yield another symmetric and periodic matrix, $\Lambda(t)$ (see
Eq.~(\ref{def_Lambda})), which gives the various covariances of the 
fluctuations in the space of the Floquet solutions. However, the focus of our 
interest is in the two-time correlations of the fluctuations which are shown 
in Appendix \ref{app:c} to equal
$C(t+\tau,t)=2 P(t+\tau)Y(\tau)\Lambda(t)P^\mathrm{T}(t)$. Therefore the 
autocorrelation function itself equals 
\be\label{eq:auto}
C(\tau)=\frac{2}{T}\int_0^{T}P(t+\tau)Y(\tau)\Lambda(t)P^\mathrm{T}(t)\dd t.
\ee
for $\tau\geq0$. The diagonal elements of $C(\tau)$ turn out to be even 
functions of $\tau$, as they ought to be. Power spectra, $P_i(\omega)$ for 
$i = 1,\ldots,d$, may then be calculated as the Fourier transform of diagonal 
elements of $C(\tau)$, as in Eq.~(\ref{eq:Fourier}).

\subsection{The case of two coupled Brusselator systems}
\begin{figure}\center
\includegraphics[width=.95\columnwidth]{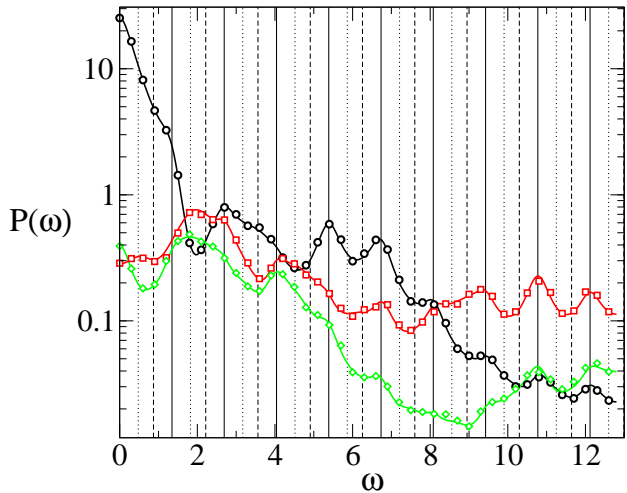}
\vspace{1em}
\caption{\label{fig:coupled} (Color online) Power spectra of the fluctuations 
about the limit cycles of the coupled Brusselator system. Data is shown for 
the three transverse directions, with black circles indicating the normal 
direction, dark gray (red) squares the binormal component, and light gray 
(green) diamonds the trinormal directions.  Markers show results from
simulations at a system size of $2\times 10^5$ and averaged over $2000$ runs. 
The solid lines are from the theory, and as seen in the figure the agreement 
with simulations is near perfect. Vertical lines are shown at frequencies 
$n\omega_0$ (solid), $n\omega_0+\mathrm{Im}\,\mu$ (dotted) and 
$n\omega_0-\mathrm{Im}\,\mu$ (dashed), with $n$ a positive integer.}
\end{figure}

In order to demonstrate the method on a concrete example, we will study a 
model composed of two coupled Brusselator systems. Two Brusselator units 
can be coupled in a number of different ways and here we construct the 
coupling in such as way as to draw parallels with the forced Brusselator 
discussed earlier. Chemical species $X_1$ and $X_2$ form a primary Brusselator
through reactions, (\ref{eq:r1})-(\ref{eq:r4}), with constant populations of 
$A$, $B$ and $C$. We now also introduce species $X_3$, $X_4$ and $C'$, which 
follow the reactions, 
\BE\label{eq:r5}
A&\rightarrow&X_3+A,\\\label{eq:r6}
X_3&\rightarrow&\emptyset,\\\label{eq:r7}
X_3+X_2&\rightarrow&X_4+X_2,\\\label{eq:r8}
2X_3+X_4+C'&\rightarrow&3X_3+C'.
\EE
Given that substance $A$ is part of both units, the secondary Brusselator 
therefore has the same system size as the primary one. The deterministic 
dynamics is given by 
\BE
\dot{x}_1&=&1-x_1(1+b-cx_1x_2),\\
\dot{x}_2&=&x_1(b-cx_1x_2),\\
\dot{x}_3&=&1-x_3(1+x_2-c'x_3x_4),\\
\dot{x}_4&=&x_3(x_2-c'x_3x_4).
\EE
When $b>1+c$ there is a limit cycle in the primary Brusselator; we will again 
denote its angular frequency by $\omega_0$. These oscillations of the primary 
Brusselator act as a periodic forcing on the second, and for all parameters 
studied here, the second Brusselator shows cycles at the above frequency 
$\omega_0$. The two Brusselators together form a four-dimensional
autonomous system. Hence, we will study the fluctuations of the large
system-size discrete system which act transverse to the limit
cycle. In this example then, we discuss the normal, $\hat{\bolde}_2$, binormal
$\hat{\bolde}_3$, and trinormal $\hat{\bolde}_4$, directions. Once the 
periodic drift $K(t)$ and diffusion $D(t)$ matrices, given by 
Eqs.~(\ref{eq:coup_K1})-(\ref{eq:coup_D2}) in Appendix \ref{app:a}, are 
rotated into the Frenet frame, we then calculate power spectra of transverse 
fluctuations via Eq.~(\ref{eq:auto}). The results of this are presented in 
Fig.~\ref{fig:coupled} for the model parameters $b=3.3$, $c=2$, and $c'=1$.
We find very good agreement between theory and simulation performed using the 
Gillespie algorithm. For these parameters, one of the non-trivial Floquet 
multipliers, $\rho_2$, is real and positive, while the remaining two, 
$\rho_\pm$, take on complex conjugate values. However, these multipliers are 
not associated with any particular transverse direction, as can be seen
from the power spectra: in all three directions, $\hat{\bolde}_2$,
$\hat{\bolde}_3$, and $\hat{\bolde}_4$, peaks are found at frequencies equal to
a multiple of $\omega_0$, but also at those associated with the imaginary 
parts of the complex Floquet exponents, $\omega_0+\mathrm{Im}\,\mu_\pm$. While 
the general formalism we have developed in this section has been illustrated 
on the concrete example of the coupled Brusselator, it should be clear that 
it can be applied quite generally to investigate the fluctuations about a 
limit cycle in $S$-dimensions. 

\section{Conclusions}
\label{conclude}
The phenomenon of stochastic amplification due to demographic, or intrinsic, 
noise has been qualitatively understood for fifty years, but it is only 
recently that it has been comprehensively and quantitatively described. This 
has been due in large part to the application of the technique of the 
system-size expansion, which is able to reproduce results obtained by 
numerical simulations to a remarkable precision. In fact, although this method 
allows for a systematic expansion in powers of $1/\sqrt{N}$, there is usually 
no need to go beyond next-to-leading order. In essence, application of the 
method means that the use of numerical simulations to understand the cycles 
induced by noise could be dispensed with entirely.

If the systems under study are subject to an external periodic driving, for 
example biological systems subject to an annual cycle, then the deterministic 
dynamics may have a limit cycle as its stable state. In this paper we have 
investigated the effect that demographic stochasticity will have on this state.
On general grounds one might expect that if the limit cycle was approached 
in an oscillatory manner, then stochastic cycles about the limit cycle could 
be sustained. We have shown that once again the system-size expansion may be 
applied to gain a quantitative understanding of this phenomena. The analysis 
is considerably more elaborate than in the case where the deterministic 
dynamics approaches a fixed point, but once again the method gives excellent 
agreement with numerical simulations.  

The signature for the oscillatory approach to limit cycles is that the 
associated Floquet multiplier should be complex. This can occur for 
nonautonomous systems in two or more dimensions or autonomous systems in three
or more dimensions. Since the eigenvalues of a typical real matrix in these 
dimensions will generically be complex, one might expect complex Floquet 
exponents to be common. Our investigations of various models, although far
from comprehensive, suggests that they are quite common in periodically driven 
systems, but not so common in autonomous systems that are generally studied. 
There may be a dynamical reason for this, but it is as likely that this is due 
to the nonlinear systems appearing in the literature being selected for their 
period doubling transition to chaos, rather for the structure of their limit 
cycles. 

In the past it was said that intrinsic noise could turn oscillatory decay to 
a fixed point into sustained oscillations. It was expected that these 
oscillations would have periods ${\rm Im}\,\lambda_i$, where $\lambda_i$ were 
the eigenvalues of the stability matrix for that fixed point. This is only 
true in a very broad sense, as studies over the last few years have shown. In 
reality the period may significantly deviate from ${\rm Im}\,\lambda_i$ due to 
other factors, and the amplitude of the fluctuations may be much larger than 
might be expected due to a resonance effect. Analogously, one might guess that 
intrinsic noise could turn oscillatory decay to a limit cycle into sustained 
oscillations about that cycle and that these oscillations would have 
periods $n\omega_0 \pm {\rm Im}\,\mu_i$, where $\omega_0$ is the period of the 
limit cycle and $\mu_i$ are the Floquet exponents associated with that 
limit cycle. We have shown in this paper that this is indeed the case in a 
broad sense, but as for the case of the fixed point there is much more to the
story than this. For instance, the expressions $n\omega_0 \pm {\rm Im}\,\mu_i$
are again just an approximation to the frequencies and the amplitude of the 
oscillations will vary significantly depending on a number of factors, such as 
the magnitude of the Floquet multipliers. Fortunately, the system-size 
expansion once again gives results which are in excellent agreement with 
simulations and gives us a way of exploring the nature of these fluctuations. 
We expect that the ideas presented in this paper will have a number of 
applications, which we hope to explore and report on in the future.

\begin{acknowledgments} 
We would like to thank D. Broomhead, R. Devaney and J. J. Tyson for useful 
discussions, which helped to shape the work presented here. This work was 
supported through a PhD Scholarship (RCUK reference EP/50158X/1) to RPB and 
an RCUK Fellowship to TG (RCUK reference EP/E500048/1).
\end{acknowledgments}

\begin{appendix}

\section{Explicit Forms of Matrices}
\label{app:a}
The matrices which appear in the description of the fluctuations about the 
deterministic trajectory are given in this appendix. The drift matrix $K(\bx)$ 
and the diffusion matrix $D(\bx)$ are naturally functions of the concentration 
$\bx$. However, when the solutions of the deterministic dynamics, 
$\bolx(t+T)=\bolx(t)$, are limit cycles they themselves become periodic 
functions of time. For the remainder of this appendix, we shall suppress the 
time dependence of $\bolx(t)$ for greater clarity.
\subsection{Forced Brusselator}
\be\label{eq:forcedK}
K(t)=\left(\begin{array}{rr}
-\left[ 1+b(t)-2c\ovx_1\ovx_2 \right] & c\ovx_1^2 \cr \cr
\left[ b(t)-2c\ovx_1\ovx_2 \right] & -c\ovx_1^2
\end{array}\right),
\ee
\be\label{eq:forcedD}
D(t)=\left(\begin{array}{rr}
D_1(t)&-D_2(t) \cr \cr
-D_2(t)&D_2(t)
\end{array}\right),
\ee
where
\BE\nonumber
D_{1}(t)&=& \frac{1}{2} \left\{ 1+\ovx_1 \left[ 1+b(t)+
c\ovx_1\ovx_2 \right] \right\} ,\\\label{apeq:D}
D_{2}(t)&=& \frac{1}{2} \left\{ \ovx_1 \left[ b(t) 
+ c\ovx_1\ovx_2 \right] \right\} .
\EE
\subsection{Willamow\-ski-R\"ossler Model}

\be\label{eq:will_K1}
K(t)=\left(\begin{array}{ccc}
K_{11}(t)  & -\ovx_{1} & -\ovx_{1}\cr \cr
-\ovx_{2} & K_{22}(t) & 0\cr \cr
\ovx_{3} & 0 & K_{33}(t)
\end{array}
\right),
\ee
where
\BE\nonumber
K_{11}(t)&=& b_{1}-2d_{1}\ovx_{1}-\ovx_{2}-\ovx_{3},\\\nonumber
K_{22}(t)&=& b_{2}-2d_{2}\ovx_{2}-\ovx_{1}, \\\label{eq:will_K2}
K_{33}(t)&=& \ovx_{1}-d_{3}, 
\EE
and
\be\label{eq:will_D}
D(t)=\left(
\begin{array}{ccc}
D_{11}(t)&D_{12}(t)&D_{13}(t)\cr \cr
D_{12}(t)&D_{22}(t)&0\cr \cr
D_{13}(t)&0&D_{33}(t)
\end{array}
\right),
\ee
where
\BE
D_{11}(t)&=& \frac{1}{2} \ovx_1(b_1 + d_1\ovx_1 + \ovx_2 + \ovx_3), 
\nonumber \\
D_{12}(t)&=& \frac{1}{2} \ovx_1\ovx_2, \nonumber \\
D_{13}(t)&=&- \frac{1}{2} \ovx_1\ovx_3, \nonumber \\
D_{22}(t)&=& \frac{1}{2} \ovx_2(b_2 + d_2\ovx_2 + \ovx_1),\nonumber \\
D_{33}(t)&=& \frac{1}{2} \ovx_3(\ovx_1 + d_3).
\EE

\subsection{Coupled Brusselators}
\be\label{eq:coup_K1}
K(t)=\left(
\begin{array}{rrrr}
K_{1}(t)-1&c\ovx_1^2&0&0\cr \cr
-K_{1}(t)&-c\ovx_1^2&0&0\cr \cr
0&-\ovx_3&K_{3}(t)-1&c'\ovx_3^2\cr \cr
0&\ovx_3&-K_{3}(t)&-c'\ovx_3^2
\end{array}
\right),
\ee
where
\BE\label{eq:coup_K2}
K_{1}(t)&=&2c\ovx_1\ovx_2-b, \nonumber \\
K_{3}(t)&=&2c'\ovx_3\ovx_4-\ovx_2,
\EE
and
\be\label{eq:coup_D1}
D(t)=\left(
\begin{array}{rrrr}
D_1(t)&-D_2(t)&0&0\cr \cr
-D_2(t)&D_2(t)&0&0\cr \cr
0&0&D_3(t)&-D_4(t)\cr \cr
0&0&-D_4(t)&D_4(t)
\end{array}
\right),
\ee
where in addition to (\ref{apeq:D}) we have,
\BE\label{eq:coup_D2}
D_{3}(t)&=& \frac{1}{2} \left[ 1+\ovx_3\left(1+\ovx_2+
c'\ovx_3\ovx_4\right) \right],
\nonumber \\
D_{4}(t)&=& \frac{1}{2}\ovx_3\left(\ovx_2+c'\ovx_3\ovx_4\right).
\EE

\section{The Frenet Frame}
\label{app:b}
In this appendix we will discuss the background, and develop the formalism, 
relating to the co-moving frame which we use to study the fluctuations from 
the limit cycles discussed in the main text. Such a frame, called a Frenet 
frame \cite{kuhnel}, is a natural way to study displacements from 
a deterministic trajectory in any number of dimensions. Here we will denote 
the number of dimensions by $S$.

Consider the general autonomous problem which we may describe by a system of 
non-linear, homogeneous, first-order equations, 
\be\label{eq:autonomous}
\frac{\dd \bx}{\dd t}=\bA(\bx).
\ee
Following \cite{kuhnel}, we may define the Frenet frame by applying the 
Gram-Schmidt orthogonalization procedure to the time-derivatives of the 
solution, $\bx(t)$. So long as the time derivatives are linearly 
independent, this gives the basis vectors, $\hat{\bolde}_i(t)$, of the frame 
to be
\BE\label{eq:frenet1a}
\bolde_i(t)&=&\frac{\dd^i{\bx}(t)}{\dd t^i}-\sum_{j=1}^{i-1}
\left( \frac{\dd^i{\bx}(t)}{\dd t^i}
\cdot\hat{\bolde}_j(t)\right)\hat{\bolde}_j(t),\\\label{eq:frenet2a}
\hat{\bolde}_i(t)&=&\frac{\bolde_i(t)}{\left| \bolde_i(t)\right|}, \ 
i\in\{1,\ldots,S\}.
\EE
We may now construct the matrix which transforms from Cartesian co-ordinates 
to the Frenet frame to be
$J(t)=\left(\hat{\bolde}_1(t),\ldots,\hat{\bolde}_S(t)\right)^\mathrm{T}$.
The transformation is by construction an orthogonal matrix $\mathrm{O}(S)$, 
and as such has the property that $J^\mathrm{T}(t)=J^{-1}(t)$ for all times. 

We now wish to consider the effect of this transformation on the equation of 
a linear fluctuation, $\bxi(t)$, about the deterministic solution, 
$\bolx(t)$. For the time being we will neglect the noise term and consider the 
homogeneous equation, $\dot{\bxi}(t)=K(t)\bxi(t)$. The transformation to the 
Frenet frame takes the form $\bxi(t)\mapsto\bq(t)=J(t)\bxi(t)$. Then 
$\dot{\bxi}(t)= ( \dot{J}(t) + J(t)K(t))\bxi(t)$ and so the rotated 
displacement obeys the linear equation,
\be\label{eq:full}
\dot{\bq}(t)=K^{\rm tot}(t)\bq(t),
\ee
where $K^{\rm tot}(t) = K'(t) + R(t)$ and where
\be 
K'(t) = J(t) K(t) J^{-1}(t), \  R(t) = \dot{J}(t) J^{-1}(t).
\label{K_tot}
\ee

We now evaluate the elements of the first column of the matrix $K^{\rm tot}$.
These have an especially simple form, with $K^{\rm tot}_{i 1} =0$ for
$i > 1$. This follows from the fact that, for an autonomous system, 
$\ddot{\bolx}(t) = K(t)\dot{\bolx}(t)$, and so the ``velocity'' 
$\dot{\bolx}(t)$, is a solution of the homogeneous equation that we are 
considering. From this, and 
from $\hat{\bolde}_1(t)=\dot{\bolx}(t)/|\dot{\bolx}(t)|$, it follows that
\be\label{eq:first}
K'_{i1}(t)=\frac{1}{|\dot{\bolx}(t)|}\hat{\bolde}_i(t)\cdot\ddot{\bolx}(t).
\ee
The second term in the definition of $K^{\rm tot}(t), R(t)$, may be written 
in terms of the basis vectors and, due to their orthogonality properties, we 
have 
\be
R_{i1}(t) = \frac{\dd\hat{\bolde}_i(t)}{\dd t}\cdot\hat{\bolde}_1(t)=
-\hat{\bolde}_i(t)\cdot\frac{\dd\hat{\bolde}_1(t)}{\dd t},
\ee
for $i \neq 1$. The rate of change of the longitudinal basis vector is given 
by $\big(\ddot{\bolx}(t)-
\hat{\bolde}_1(\hat{\bolde}_1\cdot\ddot{\bolx}(t))\big)/|\dot{\bolx}(t)|$ and 
so
\be\label{eq:second}
R_{i1}(t)=-\frac{1}{|\dot{\bolx}(t)|}\hat{\bolde}_k(t)\cdot\ddot{\bolx}(t), 
\ i \neq 1.
\ee
Adding Eqs.~(\ref{eq:first}) and (\ref{eq:second}), and noting that $R_{11}=0$,
we have
\be
K^{\rm tot}_{i1}(t) = 0\ (i>1); \ K^{\rm tot}_{11}(t) 
=\frac{1}{|\dot{\bolx}(t)|^2}\dot{\bolx}(t)\cdot\ddot{\bolx}(t).
\label{K_i1}
\ee
So all of the elements of the first column of $K^{\rm tot}(t)$ vanish, apart 
from the element which is also in the first row. The significance of this is 
that the transverse displacements, which we denote by $\mathbf{r}(t)$ 
decouple from the longitudinal displacements, denoted by $s(t)$. So writing
a general displacement as $\bq(t)=(s(t),\mathbf{r}(t))$, we have
\BE
\label{longit}
\dot{s}(t)&=&K^{\rm tot}_{11}(t)s(t)+\mathbf{K}_{sr}(t)\cdot\mathbf{r}(t),\\
\dot{\mathbf{r}}(t)&=&\widetilde{K}(t) \mathbf{r}(t),
\label{trans}
\EE
where the vector $\mathbf{K}_{sr}(t)$ is the $(S-1)$-dimensional vector 
$K^{\rm tot}_{1i}(t)$ and where $\widetilde{K}(t)$ now describes the purely 
transverse drift behavior. So the Frenet frame always separates the equation 
of motion for the linear fluctuations into longitudinal and transverse parts 
and the transverse motion is free from any influence by the longitudinal 
motion.

\section{Autocorrelations of Periodic Langevin Equations}
\label{app:c}
The equations which describe small perturbations about the limit cycle 
either have the form (\ref{eq:linear}) for non-autonomous (forced) systems
or the form (\ref{trans}) for autonomous (unforced) systems. In the latter 
case longitudinal displacements have been excluded, but once this has been 
done, the analysis for both cases is identical. So we can develop the theory 
for both together, we will adopt the notation of the autonomous case, that is, 
start from the equation
$\dot{\mathbf{r}}(t)=\widetilde{K}(t)\mathbf{r}(t)$. It should then be
understood that in the non-autonomous case the replacements
$\mathbf{r}(t) \to \bxi(t)$ and $\widetilde{K}(t) \to K(t)$ should be made.

The results of Floquet theory \cite{grimshaw} tell us that, when 
$\widetilde{K}(t+T)=\widetilde{K}(t)$ for all $t$, one may generally find 
$d$ linearly independent solutions to the homogeneous equation 
$\dot{\mathbf{r}}(t)=\widetilde{K}(t)\mathbf{r}(t)$ which have the form 
$\mathbf{r}(t)=\bp_i(t)\mathrm{e}^{\mu_i t}$. Here $\mu_i$, $i = 1,\ldots,d$,
are the Floquet exponents, which may in general be complex, and the 
functions $\bp_i(t)$ are periodic with the period, $T$. From these solutions,
the the canonical fundamental matrix, $X(t)$, may be constructed. It has the 
special property that the constant Floquet matrix, $B=X^{-1}(t)X(t+T)$, is 
diagonal with elements equal to the Floquet multipliers. 
Grimshaw \cite{grimshaw} appends a subscript $0$ to denote the canonical 
choice which results in a diagonal Floquet matrix, but since we will only deal 
with such a choice in this paper, we omit this subscript. However when carrying
out numerical work, it should be recognized that in general the solutions 
which are found will be linear combinations of solutions of the form 
$\bp_i(t)\mathrm{e}^{\mu_i t}$. These can be used to find a (non-diagonal) 
$B$, the eigenvectors of which can be used to construct a similarity 
transformation to a canonical form. An alternative way of describing the 
canonical solutions is to define the periodic matrix 
$P(t)=(\bp_1(t),\ldots,\bp_d(t))$ and the diagonal exponential matrix 
$Y(t)=\exp\{\mathrm{diag}(\mu_1\ldots\mu_d)t\}$. In terms of these the 
canonical fundamental matrix is given by $X(t)=P(t)Y(t)$.

Moving on to the fluctuations about the periodic solutions of the 
deterministic dynamics, the linear stochastic fluctuations obey a Langevin 
equation (\ref{eq:lang_forcedbruss}), with the noise correlator given by
Eq.~(\ref{eq:matrixd}), for non-autonomous (forced) systems and a 
Langevin equation (\ref{eq:rot_lang}), with the noise correlator given by
Eq.~(\ref{eq:matrixg}), where $G(t)=J(t)D(t)J^{-1}(t)$, for autonomous 
(unforced) systems. To separate out the latter into longitudinal and transverse
components, we note that in Appendix \ref{app:b} we wrote
$\bq(t)=(s(t),\mathbf{r}(t))$, and now we analogously write  
$\bg(t)=(g_s(t),\bg_r(t))$. Then, since the transverse fluctuations decouple
from the longitudinal fluctuations, the Langevin equation for purely transverse
fluctuations $\mathbf{r}(t)$ may be written as
\be\label{eq:transverse_Lang}
\dot{\mathbf{r}}(t)=\widetilde{K}(t)\mathbf{r}(t) + \bg_r(t).
\ee 
The noise correlator (\ref{eq:matrixg}) can be expressed in terms of
transverse and longitudinal components by decomposing $G(t)$ as follows:
\be\label{eq:allG}
G(t)=\left(
\begin{array}{cc}
G_{ss}(t)&\mathbf{G}^\mathrm{T}_{sr}(t)\\
\mathbf{G}_{sr}(t)&\widetilde{G}(t)
\end{array}
\right).
\ee
Since the vector $\mathbf{G}_{sr}(t)$ is typically non-zero, the random 
variables, $g_s$ and $\bg_r$, generally remain statistically correlated in 
the rotated frame. However, this is only important if we intend to evaluate 
simultaneous values of both $g_s(t)$ and $\bg_r(t)$ and this we do not do, 
because we have already shown for the noiseless case that the transverse 
displacements are independent of longitudinal one. Therefore the only noise 
correlator we require is
\be\label{eq:corr_trans}
\avg{\bg_r(t)\cdot\bg_r^\mathrm{T}(t')}= 2 \widetilde{G}(t)\delta(t-t').
\ee
Once again we will develop the theory using the notation of 
Eqs.~(\ref{eq:transverse_Lang}) and (\ref{eq:corr_trans}), but it applies 
equally to Eqs.~(\ref{eq:lang_forcedbruss}) and (\ref{eq:matrixd}).

Floquet theory may be applied to linear inhomogeneous equations of the form 
(\ref{eq:transverse_Lang}), as well as to homogeneous equations such as
$\dot{\mathbf{r}}(t)=\widetilde{K}(t)\mathbf{r}(t)$ \cite{grimshaw}. To 
solve Eq.~(\ref{eq:transverse_Lang}), we proceed in the standard way and add 
a particular solution of the equation to a general solution of the 
corresponding homogeneous equation. This yields \cite{grimshaw}
\be\label{eq:linLang_solution}
\mathbf{r}(t)=X(t)\mathbf{r}_0+X(t)\int_{t_0}^tX^{-1}(s)\bg_r(s)\dd s,
\ee
for $t\geq t_0$ and with the initial condition 
$\mathbf{r}(t_0)= X(t_0)\mathbf{r}_0$. Since we will not be interested in the 
effects of transients in this paper, we set the initial conditions in the 
infinitely distant past, $t_0\to-\infty$. A change of integration variable 
$s\to s'=t-s$ in the solution (\ref{eq:linLang_solution}) now gives 
\be\label{eq:stationary_solution}
\mathbf{r}(t)=P(t)\int_0^\infty Y(s')P^{-1}(t-s')\bg_r(t-s')\dd s',
\ee
where we have used the fact that, since $Y(t)$ is a diagonal exponential 
matrix, $Y(t_1 + t_2)=Y(t_1)Y(t_2)$.

Of course, $\mathbf{r}(t)$ is a stochastic variable, and we will typically be
interested in finding correlation functions, principally the two-time 
correlation function 
$C(t+\tau,t)=\avg{\mathbf{r}(t+\tau)\mathbf{r}^{\mathrm{T}}(t)}$. Taking
$\tau\geq0$, the solution (\ref{eq:stationary_solution}) gives
\be\label{eq:two-time}
C(t+\tau,t)= 2 P(t+\tau)Y(\tau)\Lambda(t)P^\mathrm{T}(t),
\ee
where we have introduced the symmetric and periodic matrix integral,
\be
\label{def_Lambda}
\Lambda(t)=\int_0^\infty Y(s)\Gamma(t-s)Y(s)\dd s,
\ee
and, in turn, the symmetric and periodic matrix
\be
\label{def_Gamma}
\Gamma(s)=P^{-1}(s)\widetilde{G}(s)\left(P^{-1}(s)\right)^\mathrm{T}.
\ee
All of the functions in Eq.~(\ref{eq:two-time}) are deterministic and may be 
evaluated given a good numerical estimate for the limit cycle solution 
$\bolx(t)$. 

The infinite integral for $\Lambda(t)$ may be evaluated as a re-summed finite 
integral due to the periodicity of $\Gamma(s)$. The result, in terms of 
Floquet multipliers, $\rho_i$, is then,
\be\label{eq:Lambda}
\Lambda_{ij}(t)=\frac{1}{1-\rho_i\rho_j}\int_0^{T}\mathrm{e}^{(\mu_i+\mu_j)s}
\Gamma_{ij}(t-s)\dd s,
\ee
for $i,j = 1,\dots,d$. The origin of the prefactor is from an infinite 
geometric summation, $\sum_{n=0}^\infty(\rho_i\rho_j)^n$, which is convergent 
when the Floquet multipliers are inside the unit circle.

Finally, although the details are not presented here, an expression can be 
found for  $\tau<0$. It turns out that $C(\tau) = C(-\tau)^\mathrm{T}$, as it 
ought. Hence the final form is given by Eq.~(\ref{eq:auto}) for $\tau \geq 0$, 
and can be found from  Eq.~(\ref{eq:auto}) for $\tau \leq 0$, supplemented by
the condition $C(\tau) = C(-\tau)^\mathrm{T}$.

\end{appendix}
\end{document}